\documentstyle[12pt]{article}
\newcommand{\nc}{\newcommand}
\nc{\beq}{\begin{equation}} \nc{\eeq}{\end{equation}}
\nc{\beqa}{\begin{eqnarray}} \nc{\eeqa}{\end{eqnarray}}
\nc{\lsim}{\begin{array}{c}\,\sim\vspace{-21pt}\\< \end{array}}
\nc{\gsim}{\begin{array}{c}\sim\vspace{-21pt}\\> \end{array}}

\nc{\scR}{{\cal R}}
\nc{\scL}{{\cal L}}
\nc{\al}{\alpha}
\nc{\ald}{\dot{\alpha}}
\nc{\be}{\beta}
\nc{\bed}{\dot{\beta}}
\nc{\lam}{\lambda}
\nc{\nud}{\dot{\nu}}
\nc{\lamd}{\dot{\lam}}

\newcommand{\drawsquare}[2]{\hbox{%
\rule{#2pt}{#1pt}\hskip-#2pt%  left vertical
\rule{#1pt}{#2pt}\hskip-#1pt%  lower horizontal
\rule[#1pt]{#1pt}{#2pt}}\rule[#1pt]{#2pt}{#2pt}\hskip-#2pt%  upper horizontal
\rule{#2pt}{#1pt}}% right vertical

\newcommand{\Yfund}{\raisebox{-.5pt}{\drawsquare{6.5}{0.4}}}%  fund
\nc{\Ap}[2]{A^\prime_{#1#2}}

\nc{\Q}[2]{Q_{#1#2}}
\nc{\R}[2]{R_{#1#2}}
\nc{\Y}[2]{Y_{#1#2}}
\nc{\V}[2]{V_{#1#2}}
\nc{\q}[2]{q_{#1#2}}
\nc{\G}[2]{G_{#1#2}}
\nc{\W}[2]{W_{#1#2}}
\nc{\D}[2]{D_{#1#2}}
\nc{\A}[2]{A_{#1#2}}
\nc{\p}[2]{p_{#1#2}}
\nc{\vv}[2]{v_{#1}^{#2}}
\nc{\rr}[2]{r_{#1}^{#2}}
\nc{\lij}[2]{l_{#1}^{#2}}

\nc{\spa}{SU(2)_1}
\nc{\spb}{SU(2)_2}
\nc{\spc}{SP(2n-4)}
\nc{\spd}{SP(4n+2m-10)}
\nc{\Lsc}[2]{\scL_{#1#2}}
\nc{\Lrm}[2]{L_{#1#2}}
\nc{\Rsc}[2]{\scR_{#1#2}}

\begin{document}

\begin{titlepage}

\vspace{2cm}

{\hbox to\hsize{hep-ph/9609529 \hfill EFI-96-35}}
{\hbox to\hsize{September 1996 \hfill Fermilab-Pub-96/338-T}}
{\hbox to\hsize{
\hfill revised version }}

\bigskip

\begin{center}

\vspace{2cm}

\bigskip

\bigskip

\bigskip

{\Large \bf New Models of Gauge and Gravity 
Mediated Supersymmetry Breaking}

\bigskip

\bigskip

{\bf Erich Poppitz}$^{\bf a}$ and {\bf Sandip P. Trivedi}$^{\bf b}$ \\

\bigskip

\bigskip

$^{\bf a}${\small \it Enrico Fermi Institute\\
 University of Chicago\\
 5640 S. Ellis Avenue\\
 Chicago, IL 60637, USA\\

{\tt epoppitz@yukawa.uchicago.edu}\\}
\smallskip

 \bigskip

$^{\bf b}${ \small \it Fermi National Accelerator Laboratory\\
  P.O.Box 500\\
  Batavia, IL 60510, USA\\

 {\tt trivedi@fnal.gov}\\ }

\vspace{1.3cm}

\begin{abstract} 

We show that supersymmetry breaking in a class of theories with 
$SU(N) \times SU(N-2)$
gauge symmetry can be studied in a calculable sigma model. We use the 
sigma model to show that the supersymmetry breaking vacuum in these theories 
leaves
a large subgroup of flavor symmetries intact, and to calculate
the masses of the low-lying states.  By embedding the Standard Model 
gauge groups in the unbroken flavor symmetry group
we construct a class of models in which supersymmetry breaking is 
communicated by both gravitational and gauge 
interactions. One distinguishing feature of these models is that the 
messenger fields, responsible for the gauge mediated communication of 
supersymmetry breaking, are an integral part of the supersymmetry breaking 
sector.
We also show how, by lowering the scale that suppresses the nonrenormalizable
operators, a 
class of purely gauge mediated
models with a combined supersymmetry breaking-cum-messenger sector
can be built. We briefly
discuss the phenomenological features of the models we construct.

\end{abstract}

\end{center}

\end{titlepage}

\renewcommand{\thepage}{\arabic{page}}
\setcounter{page}{1}

\baselineskip=18pt

\section{Introduction. }

In order to be relevant to nature, supersymmetry must be spontaneously
broken.   An  attractive idea  in this regard is that the breaking occurs
nonperturbatively  \cite{witten} 
in a strongly coupled sector of the theory and is then 
communicated to 
the Standard Model fields  by some ``messenger" interaction.  One possibility
is  that the  role of the messenger is 
played by gravity---giving rise to the so-called hidden sector models (for
a review, see~\cite{nilles}). 
Another  possibility \cite{oldpapers}, \cite{nilles}, 
which has received considerable attention recently \cite{dnns}-\cite{details} 
is that the supersymmetry 
breaking  is communicated by gauge interactions---the  gauge mediated models.  

The past few years have seen some remarkable progress in the understanding
of  non-perturbative  supersymmetric gauge theories
\cite{seibergexact}, \cite{seiberg}. This progress 
has made a more thorough investigation of supersymmetry breaking possible
\cite{susybreaking}.  
We begin this paper by extending the study of supersymmetry breaking in 
a class of theories with $SU(N) \times SU(N-2)$ gauge symmetry. These theories
were first considered in ref.~\cite{we}. We use some elegant 
observations by Y. Shirman \cite{shirman} to show that 
the low-energy dynamics of  these theories can be studied in terms of a 
calculable low-energy sigma model.  We use the sigma model to  show that the 
supersymmetry breaking vacuum in these theories preserves a large group of 
flavor symmetries,  and to calculate the spectrum of 
low-energy excitations.

We then turn to model building. The models we construct have two sectors:
a supersymmetry breaking sector---consisting of 
an $SU(N) \times SU(N-2)$ theory 
mentioned above---and the usual Standard Model sector. 
The basic idea is to embed the Standard Model gauge groups in the unbroken 
flavor symmetries of the supersymmetry breaking sector. As a result, in these
models the breaking of supersymmetry can be communicated directly by the 
Standard Model gauge groups. 
This is to be contrasted
with models of gauge mediated supersymmetry breaking 
constructed elsewhere \cite{dnns},
 in which a fairly elaborate messenger sector is needed to accomplish the 
feed-down of supersymmetry breaking. 

In the models under consideration here, 
the scale of supersymmetry breaking turns out to be high, of order  
the intermediate scale, i.e., $10^{10}$ GeV. 
As a result, the  gravity mediated effects are comparable 
to the gauge mediated ones. 
The resulting phenomenology in these ``hybrid"  models is different 
from both the 
gravity and gauge mediated cases. 
Scalars acquire both universal soft masses 
due to gravity and non-universal masses due to gauge interactions, while 
gauginos receive masses only due to gauge  interactions.  
Since the scale of supersymmetry breaking is large,  the gravitino has an  
electroweak scale mass.  Finally, there is new physics in this theory, 
at about 10 TeV,  at which scale all light degrees of freedom of the 
supersymmetry breaking sector, including those carrying Standard Model 
quantum numbers, can be probed. 

The biggest drawback of these models is the following.  Since the scale of 
supersymmetry breaking is so high, one cannot, 
at least in the absence of any information regarding the higher dimensional 
operators in the K\" ahler potential,  rule out the  presence of flavor 
changing neutral currents. In this respect these models are no better than the 
usual  hidden sector models. 

The high scale  of supersymmetry breaking arises as follows.
Within the context  of the $SU(N) \times SU(N-2)$ models,   
in order to embed the Standard Model gauge groups 
in the unbroken flavor symmetries, one is  lead to consider large values of
$N$, namely $N \ge 11$.  In these theories supersymmetry breaking occurs only 
in the presence of non-renormalizable operators in the superpotential
and the dimension of these  operators grows as $N$ grows. 
On suppressing  the effects of these operators by the 
Planck scale,  one is lead to a large supersymmetry breaking scale.  

If one lowers the scale that suppresses the non-renormalizable operators,
the supersymmetry breaking scale is lowered as well.  We use the resulting
theories  to construct purely gauge mediated models with a combined 
supersymmetry breaking and messenger sector. 
 %By lowering the
%scale that suppresses the nonrenormalizable operators, we can obtain a lower scale
%of supersymmetry breaking.  With the help of these theories with a lower
%supersymmetry breaking scale,  we construct purely gauge 
%mediated models with a combined supersymmetry breaking
%and messenger sector. 
The lower scale suppressing the non-renormalizable
operators could arise due to new nonperturbative
dynamics. It could also arise if the Standard Model gauge groups are dual
to an underlying microscopic theory. We will not explicitly discuss how
this lower scale arises here.
A brief study of the phenomenology of the purely gauge mediated models 
we construct 
reveals some features which should  be more  generally true in models of this 
type. We hope to return to a detailed phenomenological study of these  models 
in the future. 

A few more comments are worth making with respect to the models considered 
here. First, from the perspective of a hidden sector theory, the hybrid models
are  examples of  theories without any fundamental gauge singlets in which 
gauginos obtain adequately big soft masses.\footnote{For an 
example of a hidden sector theory, in which supersymmetry breaking 
involves a global supersymmetric theory and a singlet,  and  yields 
reasonable gaugino masses, see ref.~\cite{nelson}.}
  
Second, one concern about constructing models in which the 
supersymmetry breaking sector carries Standard Model charges is 
that this typically leads to a loss of asymptotic freedom for the 
Standard Model gauge groups and the existence of Landau poles at fairly low energies. 
One interesting idea on how to deal  with this problem involves dualizing \cite{seiberg} 
the theory  and regarding the resulting dual theory---which is usually better behaved in 
the ultraviolet---as the
underlying microscopic theory. In the ``hybrid" models discussed here, one finds
that the Landau poles are pushed  beyond an energy  scale of  
order $10^{16}$ GeV. 
This is a sufficiently high energy scale that even without appealing to duality their
presence might not be a big concern. For example, new GUT scale physics 
(or conceivably even  string theory related physics)  could enter at this scale. 
The non-renormalizable operators, 
mentioned above, which are responsible for the large scale of supersymmetry 
breaking  in the ``hybrid models" are also responsible for pushing up the 
scale at which Landau poles appear; 
to this extent their presence is an attractive feature which one might want to retain. 

Finally, we would like to comment on  the low-energy effective theory used to study 
the breaking of supersymmetry in the $SU(N) \times  SU(N-2)$ theories.  
This effective theory arises as follows.  First, at very high energies, 
the $SU(N-2)$ group is broken, giving rise to an effective 
theory consisting of  some moduli fields and  a pure $SU(N)$ theory coupled to a
dilaton. The $SU(N)$ theory then confines at an intermediate energy scale giving 
rise to  a  low-energy theory involving just the dilaton  and the moduli. 
Gaugino condensation in the $SU(N)$ theory gives rise to a term in the
superpotential of this low-energy theory and as  a result,  the superpotential 
has a runaway behavior  characteristic of a theory containing a dilaton.  
 However, one finds that this runaway behavior is stabilized due to a
non-trivial K\" ahler potential involving the dilaton.
It has been suggested that a similar phenomenon might be responsible for
stabilizing the runaway behavior of the dilaton in string theory \cite{sbandd}.
In the globally supersymmetric models considered here the stabilization 
occurs due to a calculable non-trivial K\" ahler potential in 
the effective theory 
linking the dilaton with the other moduli.

\section{The Supersymmetry Breaking Sector.}

\subsection{The ${\bf SU(N)\times SU(N-2)}$ Models.}
 
In this section we will briefly review the models, introduced in 
 \cite{we}, that will play the role of a supersymmetry breaking sector.
They have an $SU(N) \times SU(N-2)$ gauge group, 
with {\it odd} $N$,
and matter content consisting of a single field,
$Q_{\alpha {\dot{\alpha}}}$, that transforms as (\Yfund , \Yfund ) under the
gauge groups, $N-2$ fields, $\bar{L}^\alpha_I$, transforming as
$(\overline{\Yfund}, {\bf 1})$, and $N$ fields,  $\bar{R}^{\ald}_A$,
that  transform as  $({\bf 1}, \overline{\Yfund})$.  Here, as in the
subsequent discussion,  we denote the gauge indices of  $SU(N)$
and $SU(N-2)$  by $\alpha$ and $\ald$, respectively, while
$I = 1\ldots N-2$ and $A = 1 \ldots N$ are  flavor indices. We note that
these theories  are  chiral---no mass terms can be added for any of the matter
fields. 

We  begin  by considering the classical moduli
space. It is described by the gauge invariant mesons and baryons:
\beqa
\label{defbaryon}
Y_{IA} &=& \bar{L}_I \cdot Q \cdot \bar{R}_A~, \nonumber \\
b^{A B} &=& {1\over (N-2)!}~
\varepsilon^{A B A_1 \cdots A_{N-2} }~
\varepsilon_{\dot{\alpha}_1 \cdots \dot{\alpha}_{N-2}}
~ \bar{R}_{A_1}^{\dot{\alpha}_1} \cdots \bar{R}_{A_{N-2}}^{\dot{\alpha}_{N-2}}  ~,
\eeqa
and $\bar{\cal{B}} = Q^{N - 2} \cdot \bar{L}^{N - 2}$. These invariants are not independent
but subject to classical constraints \cite{we}.

We will consider the theory with 
the tree-level superpotential 
\beq
\label{wtreebaryon}
W_{tree} = \lambda^{IA} ~ Y_{IA} + {1 \over M^{N-5}} ~ \alpha_{AB} ~ b^{AB} ~.
\eeq
The superpotential $W_{tree}$ 
lifts all classical flat directions, provided that 
$\lambda_{IA}$ has  maximal rank,  $N-2$, 
the matrix $\alpha_{AB}$ also has  maximal 
rank ($N-1$), 
and  its cokernel 
contains the cokernel of $\lambda^{IA}$ (rank $\lambda = N - 2$). 
With this choice of couplings, $W_{tree}$ also 
preserves a nonanomalous,
flavor dependent  $R$ symmetry. To see this,  choose
for example $\alpha^{AN} = 0, \lambda^{I N} = \lambda^{I (N-1)} = 0$ 
(to lift the classical flat 
directions). Then 
one sees that
the field $\bar{R}_N$ appears in each of the baryonic terms of the superpotential
 (\ref{wtreebaryon}), while it does not appear in any of the Yukawa terms.
Assigning different $R$ charges to the four types of fields, $\bar{R}_N$,
$\bar{R}_{A < N}$, $Q$, and $\bar{L}_I$, one has to satisfy four conditions:
two conditions ensuring that the superpotential (\ref{wtreebaryon}) has 
$R$ charge 2, and
two conditions that the gauge anomalies of this $R$ symmetry vanish. 
It is easy to see that
there is a unique solution to these four conditions. 

The couplings in the superpotential will be chosen to  preserve
a maximal global symmetry\footnote{This choice of couplings, which preserves 
the maximal global symmetries, has been made for simplicity.
 For the discussion of model building that follows, it is enough to preserve an 
$SU(3) \times SU(2) \times U(1)$ symmetry. Doing so introduces extra parameters
in the superpotential eq.~(\ref{wtreebaryon}) but does not alter the 
subsequent discussion in any significant way.}. 
We will take the nonvanishing components of the Yukawa matrix to be 
$\lambda^{IA} = \delta^{IA} \lambda$, for 
$A = 1,...,N-2$. 
The antisymmetric matrix $\alpha_{AB}$ will have the following nonvanishing
elements:
 $\alpha_{AB} = a  J_{AB}$, for $A,B < N-2$ and $\alpha_{AB} =  J_{AB}$, 
for $A,B =N-1, N-2$.
This choice of couplings preserves an $SP(N-3)$ global nonanomalous 
symmetry.\footnote{In our notation $SP(2k)$ is the 
rank $k$ unitary symplectic group with $2k$ dimensional
fundamental representation. $J_{AB}$ is the $SP(2k)$ invariant tensor; we take 
$J_{12} = 1$ and $J^{AB} J_{BC} = - \delta^A_C$.} 

The dynamics of these models was discussed in \cite{we}, where it was shown that
when the superpotential (\ref{wtreebaryon})  is added, the ground state dynamically
breaks supersymmetry.
In the next section we will   study  supersymmetry breaking  in these
theories in more detail.

\subsection{The Low-Energy Nonlinear Sigma Model.}

\subsubsection{The Essential Ideas.}

We  show  in this section that  for a region of parameter space the 
breaking of supersymmetry in the   $SU(N) \times SU(N-2)$ theories  can be 
conveniently studied in a low-energy  effective theory.  We identify the 
 degrees of  
  freedom, which appear  in this  supersymmetric nonlinear sigma model, 
and  show that both  the
superpotential and the K\" ahler potential  in the sigma model  can be reliably
calculated in the region of moduli space where the vacuum is expected to occur. 
This is   interesting  since the underlying  theory that gives  rise to the sigma
model is not   weakly coupled. 
  In the following section,  we   then  
explicitly construct  and  minimize the potential responsible for  supersymmetry
 breaking, 
thereby   deducing  the unbroken flavor symmetries and    
 the  spectrum of the low-energy excitations. 

It is convenient to begin  by considering the limit
$M \rightarrow \infty$.  In this limit, the baryonic flat directions,
described by the gauge invariant fields $b^{AB}$,  are not lifted and 
the model has  runaway directions along which the energy goes to zero
asymptotically \cite{shirman}.    
 As  was mentioned above,  we take  $\lambda^{IA}$ of eq. ~(\ref{wtreebaryon}) 
to be  $\lambda^{IA} = \delta^{IA} \lambda$, for  $A = 1,...,N-2$.  
The runaway  directions  are specified 
by the  condition that  $b^{N N-1} \rightarrow \infty$ .  The other baryons 
$b^{AB}$ can  in addition be  non-zero along these directions.   We will see  that 
once one is sufficiently far  along these directions  the  low-energy dynamics 
can be described by a calculable  effective theory. 

Let us first consider the  simplest runaway direction, 
$b^{N N-1} \rightarrow \infty$, with all the other $b^{AB} =0$.
 Along this direction the  $\bar{R}$ fields
have vacuum expectation values given by   
$\bar{R}_A^{\dot{\alpha}} = v \delta_A^{\dot{\alpha}}$ with $v \rightarrow  \infty$. 
Since the  
$SU(N-2) $ symmetry  is completely broken  at a scale $v$, its
gauge bosons   get heavy  and can be intergated out.  
In the process, several components of the $\bar{R}_A$ fields  get
heavy or eaten and   can be removed from  the low-energy theory
as  well.  In
addition, on account of   the Yukawa coupling in (\ref{wtreebaryon}) all $N-2$
flavors of $SU(N)$ quarks become massive, with mass $\lambda v$,  and can be
integrated out.   Thus one is  left with an intermediate scale  effective theory 
containing the light  components  of  the $\bar{R}_A$ fields  and the 
pure $SU(N)$ gauge theory. 
  There is one slightly novel feature about the  $SU(N)$ 
group in this  effective  theory: its strong  coupling scale  $\Lambda_{1L}$  is field
dependent.  On integrating out  the $Q$ and $L$ fields  one finds that 
\beq
\label{sllow} 
\Lambda_{1L}^{3N} = \Lambda_1^{2N+2} ~ \lambda^{N-2}  b^{N ~ N-1},
\eeq
 with $ \Lambda_1$ being the scale of the ultraviolet
$SU(N)$ theory. Thus the field $b^{N ~ N-1}$ acts as a dilaton for the 
$SU(N)$ group in the low-energy  theory.   Going further down in energy one
finds that the   $SU(N)$ group confines  at a scale  $\Lambda_{1L}$, leaving  the
dilaton,  $b^{N N-1}$,  and the other  light  components of $\bar{R}_A$
as the  excitations  in   the final low-energy theory. 
Gaugino condensation in the $SU(N)$ theory gives rise to a superpotential \cite{seibergexact} 
of the form:
\beq
\label{sabaryonw}
W = \lambda^{N-2 \over N} ~\Lambda_1^{2 N + 2 \over N} 
~\left( b^{N-1 ~N} \right)^{1\over N}  
\eeq
in this low-energy theory.  

So far  we have considered the simplest runaway direction, 
$b^{N N-1} \rightarrow
\infty$, with all the other $b^{AB} =0$. There are other runaway directions,
along which some of the other  baryons   go to infinity  as  well, at a rate
comparable or faster than $b^{N N-1}$.   
In these cases  the  underlying dynamics giving rise to the effective theory 
can be sometimes different from
that  described above. 
However, one can show that  the effective theory, 
consisting of the light 
components of  $\bar{R}_A$,  with the non-perturbative superpotential  
(\ref{sabaryonw}),    describes the low-energy dynamics  along these 
directions as well. 

It is not  surprising that the  exact superpotential can be  calculated in this 
effective theory. 
 What is more remarkable is that, as  has been
argued in  \cite{shirman},   the  corrections to the classical  K\" ahler potential are
small along these runaway directions and thus the K\" ahler potential can be
calculated in the  effective theory as well.   
Thus, as promised above, the effective theory  is completely calculable.  
Let us briefly  summarize Shirman's argument here. 
Since the $SU(N-2)$
theory is  broken at a high scale,
the corrections to the K\" ahler potential one is worried 
about must involve the  effects  of  the strongly coupled $SU(N)$ 
group\footnote{As mentioned above along some of the 
runaway directions the  underlying 
dynamics is somewhat different.  Correspondingly the strongly coupled effects
do not always involve the full $SU(N)$ group. However, an analogous argument
shows that the corrections to the classical K\" ahler potential are  small along these 
directions as well.} 
with a strong coupling scale $\Lambda_{1L}$,  eq.~(\ref{sllow}).   These 
corrections  are  
 of the form $\bar{R}^\dagger \bar{R} f(t)$, with  $t = \Lambda_{1L}^\dagger  
\Lambda_{1L}/(\bar{R}^\dagger \bar{R}) \sim 
(\Lambda_{1}^\dagger ~ \Lambda_{1})^{2 N+2  \over 3N}/(\bar{R}^\dagger
\bar{R})^{1 - (N - 2)/(3 N)}$.  We are interested  in the behavior of $f(t)$ when  $R
\rightarrow \infty$, i.e., $t \rightarrow 0$.   Now, it is easy to see that this limit 
can also be obtained when   $\Lambda_1 \rightarrow 0$.   In this case it is
clear that the strong coupling effects due to the $SU(N)$ group must  go to zero  and
thus the corrections to  the K\" ahler potential for $\bar{R}$ must be  small.  
Hereafter, we  will take the  K\" ahler potential to be classical. The discussion above 
shows that this is a good approximation as long as  $\Lambda_{1L} \ll v$,
where $v$ denotes the vacuum expectation value of the $\bar{R}$ fields. 

Let us  now briefly summarize what has been learned about the  theory when 
$M  \rightarrow \infty$.  We found that the theory had runaway directions.
The low-energy dynamics along these directions can be described by an effective 
theory consisting of the light components of the fields  $\bar{R}_A$.  Finally,
both the superpotential and the K\" ahler potential  in this effective theory can be 
calculated. 

Armed with this knowledge of the $M  \rightarrow \infty$ limit  we ask what
happens when  we consider  $M$ to be large but not infinite.  It was shown in \cite{we}
that  once the last term in  (\ref{wtreebaryon})  is turned on, the theory 
does not  have  any runaway directions and breaks supersymmetry.
 However,  and this is
the  crucial argument,   for a large enough value of $M$   the
resulting vacuum  must lie along the runaway directions discussed above
(since the runaway behavior is ultimately stopped by 
the $1/M^{N-5}$ terms in  (\ref{wtreebaryon})), and
therefore   the breaking  of supersymmetry  can be analyzed in terms of  the low-energy
theory discussed above.

\subsubsection{The Explicit Construction.} 

We now turn to explicitly constructing the  low-energy effective theory. 
The light degrees of freedom of the $\bar{R}$ fields can be described
either in terms of the appropriate components of  $\bar{R}_A$
or the gauge invariant baryons  $b^{AB}$.
The use of the baryons is more convenient \cite{ads}, 
since it automatically takes care of integrating out the 
heavy $SU(N-2)$ vector fields and their superpartners at tree level 
(see also  \cite{BPR}, \cite{RP}), and provides an explicitly
gauge invariant description of the  low-energy physics.

The K\" ahler potential for the light fields is 
$K = \bar{R}^\dagger e^{V} \bar{R} \big\vert_{V = V(\bar{R}^\dagger, \bar{R})}$, 
where the heavy vector superfield $V$ is integrated out by solving its classical equation
of motion.
In terms of the baryons, this   K\" ahler potential can
be calculated, as in \cite{ads}:
\beq
\label{Kbaryon1}
K = c_N ~ \left( ~b^\dagger_{A B} ~b^{A B} ~\right)^{1 \over N-2}  ~,
\eeq
where $c_N = (N-2) ~ 2^{-{1 \over N - 2}}$.
The baryons $b^{AB}$ are not independent,
but obey the constraints:
\beq
\label{baryonconstraints}
b^{A~B}~ b^{ N-1 ~ N} = b^{N-1 ~A} ~b^{N ~B} - b^{N -1~B} ~b^{N ~A} ~,
\eeq
which follow from their definition (\ref{defbaryon}) and Bose symmetry.
We  can now use these constraints to solve for the redundant baryons 
 in terms of    an   appropriately chosen  set  thereby 
obtaining   the required  K\" ahler potential.
Counting the number of eaten degrees 
of freedom and comparing with the analysis in terms of the fields $\bar{R}_A$
along the $D$-flat directions, it is easy to 
 see
 that $b^{N-1~ N}, ~b^{N-1~ A}$, and $b^{N~ B}$, with 
$ A,B =1,...,N-2$, are good  coordinates\footnote{For example, 
along the flat direction $\bar{R}_A^{\dot{\alpha}} = v \delta_A^{\dot{\alpha}}$,
discussed in Section 2.2.1, the components of $\bar{R}_A$ that remain light are
$\bar{R}_{N}^{\dot{\alpha}}$, $\bar{R}_{N-1}^{\dot{\alpha}}$, and $v$ (which 
describes fluctuations 
corresponding to motion along the flat direction). Using the definitions of the 
baryons (\ref{defbaryon}) one can see that fluctuations of $b^{N-1~ A}$, with 
$A < N-1$,
around these expectation values correspond to the field $\bar{R}_{N}^{\dot{\alpha}}$, 
while fluctuations of $b^{N~ A}$ ($A < N-1$) and $b^{N-1~ N}$
correspond to  $\bar{R}_{N-1}^{\dot{\alpha}}$ and $v$, respectively.}   
 (in a vacuum where $b^{N ~ N-1} \ne 0$)
and we  consequently  use them  as the  independent fields.

 For notational convenience, we  introduce the fields $S$ and 
$P^{\alpha A}$, (hereafter $A,B =1,...,N-2; \alpha = 1,2$) via the definitions:
$S = b^{N-1~N} $, $P^{1 A} = b^{N-1 A}$ and  $P^{2 A} = b^{N A} $.
The  K\" ahler  potential  (\ref{Kbaryon1}) and superpotential of the
effective theory, after using the constraint (\ref{baryonconstraints}) 
to solve for the redundant degrees of freedom,  become:
\beq
\label{effectivelagrangiankah}
K =  (N - 2)~ \left(  S^\dagger~ S +  P_{\alpha A}^\dagger ~P^{\alpha A}
+  {P^{\alpha A} ~P_{\alpha}^B~ P^\dagger_{\beta A} ~P^{\dagger \beta}_B
\over 2~ S^\dagger ~S} \right)^{1\over N-2} ,
\eeq
and 
\beq
\label{elsup}
W = \lambda^{N-2 \over N} ~\Lambda_1^{2 N + 2 \over N} ~S^{1\over N} 
- {2 \over M^{N-5}}~ P^{1~N-2} + 
{2~a \over  M^{N-5}} \sum\limits_{A, B = 1}^{N-3}  ~  
{ J_{A B}~ P^{1 A} ~P^{2 B}\over S}~,
\eeq
respectively. 
The superpotential above was obtained by adding the last term  of 
(\ref{wtreebaryon})---with the matrix $\alpha_{AB}$ chosen to 
preserve $SP(N-3)$, as described in Section 2.1---to the nonperturbatively
generated superpotential, eq.~(\ref{sabaryonw}).

We will see, in the following sections, that the sigma model has
a stable supersymmetry breaking vacuum. As discussed above, the field
$S$ is a dilaton for the $SU(N)$ gauge group. The first term in the
superpotential (\ref{elsup}) could have lead to runaway behavior. This
runaway behavior is, however, stopped by the K\" ahler potential 
(\ref{effectivelagrangiankah}), which links the dilaton to the other moduli.

\subsection{Mass Scales and Spectrum.}
\subsubsection{Mass Scales.}
With the  sigma model in hand  we can now  
write down the  the potential---it is given  in terms of the K\" ahler potential 
and the superpotential  as  $V = W_i K^{-1~i j^*} W^{*}_{j^*}$ \cite{WB}. 
The explicit minimization  of the potential in our case  needs to be done
numerically but several  features about the  resulting  ground state can be deduced 
in a straightforward way. 

Notice first, that the superpotential has two scales $\Lambda_1$ and 
$M$. These will determine   the  various scales  which  appear in this problem.  The 
scale of the vacuum expectation values $v$ can be obtained by balancing 
the first two terms in  the superpotential (\ref{elsup})  and is given  by 
\beq
\label{vevscale}
v \equiv  M ~\left[ {\lambda^{N-2 \over  2N+2} 
~\Lambda_1  \over  M  } \right]^{2N+2 \over (N-1) ~ (N-2)}~.
\eeq

In order for our approximations to be justified  $v$ needs to be large enough.
Quantitatively, we need $\Lambda_{1L}/v \ll 1$, where $\Lambda_{1L}$ is the 
strong coupling scale of the intermediate scale $SU(N)$ theory.
Since  the first term in the superpotential  (\ref{elsup}) 
is  of order $\Lambda_{1L}^3$,  
we need the condition
\beq
\label{approx}
{\Lambda_{1L} \over v} \sim  \left({v \over M}\right)^{{N-5 \over 3}} \ll 1
\eeq
to be valid. Eq.~(\ref{approx})
 can be met, for $N>5$, if $v \ll M$ \footnote{For $N = 5$, this
condition can be met  by making  a  dimensionless Yukawa coupling small.}.  
%In our discussion of model building, it will be natural to take $M = M_{Pl}$; 
% the scale  $v$ will then   turn  out to be  $\ll M_{Pl}$ so that (\ref{approx})
% will be automatically met. 
 
Hereafter it will be 
convenient to use $v$ and $M$  as the two independent  energy scales.   
The  scale of the typical  $F$ components  that give rise to 
supersymmetry breaking  is  $\sim W/v$,  i.e. 
of order $F$ where 
\beq
\label{ftermscale}
F \equiv M^2 ~ \left( {v \over M} \right)^{N - 3}~,
\eeq
while  the  masses of the fields in the sigma model  are  $ \sim W/v^2$,
i.e. of order $m$,  where
\beq
\label{orderofmasses}
m \equiv M ~ \left( {v \over M} \right)^{N - 4} ~.
\eeq
 Note  that (for $N>5$) 
the scale of supersymmetry breaking,  $F^{1/2}$, eq.~(\ref{ftermscale}), is much
higher than the scale of the masses, eq.~(\ref{orderofmasses}), if  $M \gg v$. 

We turn  now to the global symmetries. As  is clear 
from eq.~(\ref{elsup}), the superpotential has an
 $SP(N-3)$ symmetry under which 
$P^{1A}$ and $P^{2A}$ transform as fundamentals.   
First we note that although there might exist  vacua
that break the $SP(N-3)$ global symmetry,  an $SP(N-7)$ 
global symmetry is always preserved, since the light spectrum only has
two fundamentals of the $SP(N-3)$ global symmetry. 
Second, intuitively it is 
clear that  when the parameter $a$ that appears in the third term of the
superpotential (\ref{elsup})  is large, the ground state of this theory should
 preserve  the
global $SP(N-3)$ symmetry.
In the limit of large $a$, the fields that transform under the $SP(N-3)$ symmetry 
can be integrated out if the field $S$ has  an expectation value. 
The resulting theory of the light
fields  (the fields $S$ and $P^{\alpha N-2}$)  
is expected to have a stable vacuum at nonvanishing value 
of $S$ since the potential is singular for both zero and infinite field values.  
In fact,  the numerical minimization of the potential 
shows that  an   $SP(N-3)$ symmetric
stable vacuum exists for a wide range of values of $a$ (not necessarily $\gg 1$).

\subsubsection{Mass Spectrum.}

\begin{table}
{Table~1: Vacuum expectation values $x,z$, eq.~(\ref{vevs}), 
vacuum energy $\varepsilon$, and
mass matrix  
parameters $\alpha, \beta, \gamma, \delta$, 
eq.~(\ref{diracmass},\ref{scalarmass}), in the $SU(N)\times SU(N-2)$
models, for $5 \le N \le 27$, $N$-odd.}
\begin{center}
 
\vspace{0.2cm}
\label{tab1}
 
\begin{tabular}{c|c|c|c|c|c|c|c} \hline\hline
$N$ &   $x$   & $z$     & $\alpha$& $\beta$ & $\gamma$& $\delta$ & $\varepsilon$     \\
\hline
 & & & & & &\\
$5$ & .0299   & .0429   & .873    & -.437   &  .748   & .250     &   .201  \\
$7$ & .0298   & .0357   & .197    & -.0493  &  .317   & .0947    &   .0774 \\
$9$ & .0283   & .0317   & .0893   & -.0149  &  .209   & .0496    &   .0443 \\
$11$& .0262   &.0284    &.0520    &-.00650  & .159    &.0308     & .0293   \\
$13$& .0241   &.0256    &.0343    & -.00343 & .129    &.0212     &.0211    \\
$15$& .0222   & .0233   &.0245    &-.00204  &.109     &.0154     & .0159   \\
$17$& .0205   &.0214    & .0183   &-.00131  & .0946   & .0118    &.0125    \\
$19$& .0190   &.0197    &.0143    &-.000892 &.0835    &.00928    & .0100   \\
$21$& .0177   &.0183    &.0114    & -.000635&  .0748  & .00751   &.00828   \\
$23$& .0165   &.0170    &.00936   & -.000468& .0677   &.00619    & .00694  \\
$25$& .0155   &.0159    & .00780  &-.000355 & .0619   &.00520    & .00590  \\
$27$& .0146   &.0150    &.00661   &-.000275 & .0570   & .00442   &.00508   \\
 \hline\hline
\end{tabular}
\end{center}
\end{table}
%

%%%%%%%%%%%%%%%%%%%%%%%%%%%%%%%%

With this background in mind we  turn to   the    
numerical minimization.  
We will study the vacuum that preserves the maximal global symmetry 
and will in particular be interested in the masses of the 
$SP(N-3)$ fundamentals $P^{\alpha A}, A < N-2$, since they will play the role  of 
messenger fields  in the subsequent discussion of  model building. 
  The  numerical  investigation shows that  an extremum  exists 
where  the only nonvanishing vacuum expectation values are those of the fields $S$
and $P^{1 N-2}$. In particular
the field $P^{2 N-2}$ does not acquire an expectation value.\footnote{There may
exist other extrema of the potential  where also 
 the field $P^{2 N-2} \ne 0$. We have not studied these in any detail.}

The expectation values of the
fields $S$ and $P^{1 N-2}$ are:
\beqa
\label{vevs}
S &=& x ~ v^{N-2} ~, \nonumber \\
P^{1 N-2} &=& z ~ v^{N-2} ~.
\eeqa
All components of the $S$ and $P^{\alpha N-2}$ 
supermultiplets have mass
of order $m$, except the $R$ 
axion---which becomes 
massive due to higher dimensional operators \cite{BPR}, necessary
e.g. to cancel the cosmological constant---and the goldstino, which is a
linear combination of the $S$ and $P^{1 N-2}$ fermions. 
The fermionic components of the 
$SP(N-3)$ fundamentals $P^{\alpha A}$, $A = 1,...,N-3$ have a Dirac mass term,
which can be directly read off eq.~(\ref{elsup}) (the K\" ahler 
connection \cite{WB} does not contribute to the masses of 
the $SP(N-3)$ multiplets in the vacuum 
(\ref{vevs}))\footnote{In eqs.~(\ref{diracmass}) and (\ref{scalarmass})  
all kinetic terms have been brought to canonical form.}
\beq
\label{diracmass}
 \gamma~ a~ m~\sum\limits_{A, B = 1}^{N-3}~ P^{1 A}~P^{2 A}~J_{AB}, 
\eeq
while the quadratic terms in their scalar 
components are:
\beq
\label{scalarmass}
m^2 ~\sum\limits_{A, B = 1}^{N-3}~
(  P^{1 A}  ~ P^{\dagger}_{2 B} )~ 
\left( ~\begin{array}{cc} 
( \alpha + \gamma^2 ~a^2 )~\delta_A^C & \delta ~a~J_{AD}\\
\delta ~a~J^{BC} & (\beta + \gamma^2~a^2) ~\delta^B_D 
\end{array}  ~ \right)  ~
\left(  \begin{array}{c}
P^{\dagger}_{1 C} \\ P^{2 D} \end{array} \right) ~.
\eeq
The numerical values of the vacuum expectation values $x, z$
(\ref{vevs}) and the mass matrix parameters $\alpha, \beta, \gamma$,
and $\delta$, as well as the vacuum energy $\varepsilon$ 
(defined by $V = M_{SUSY}^4 =  \varepsilon~F^2~$) are given in Table 1 for a range of
values of $N$.

A few comments are now in order:

First,  it is useful to consider the 
 messenger fields'  spectrum,  (\ref{diracmass})  and  
(\ref{scalarmass}),     in the  $a \gg 1$ limit. 
The fermion mass squared and the diagonal components of the scalar mass
 matrix become equal  in this limit.   Furthermore, 
the fermion mass squared is equal to the average of the squared masses
of the scalar mass eigenstates, and the splitting in the supermultiplet (proportional
to $\sqrt{a}$) is much smaller than the supersymmetric mass (proportional to $a$).
The  spectrum  of the messenger fields  in this limit  is very similar to  that 
obtained  in the models   of   ref.~\cite{dnns}, where gauge singlet fields are responsible
for generating both the  supersymmetric and supersymmetry breaking masses.
  This  is  because  in the $a \gg 1$ limit, the masses
of the $SP(N-3)$ fundamentals   mainly arise  due 
to the last term in the superpotential in eq.~(\ref{elsup}), 
which has the form of the   singlet---messenger fields 
coupling  in the models
of ref.~\cite{dnns}.

Second,   it is very likely---at least in the   $a \gg 1$  limit---that  the 
vacuum we have explored here is in fact the global minimum of the theory. 
This is to be contrasted with  the  models   of  ref.~\cite{dnns}, 
which contain a more elaborate messenger sector. 
 In these models,   the required vacuum---with an $F$ term  expectation value
 for the 
singlet,  which couples to   the  messenger 
quarks---is  only local.  Usually there is a deeper  minimum, in which the 
singlet $F$ term expectation value
 vanishes, while the messenger  quarks  have expectation values,
breaking the Standard Model gauge group at an  unacceptably high scale
(avoiding this problem requires an even more complicated messenger 
sector, as shown in  ref.~\cite{bogdanlisa}).   

In addition to the fields in the sigma model,  when discussing the
communication of supersymmetry breaking to the Standard Model sector,
we will need some information
on the spectrum of heavy fields in the $SU(N)\times SU(N-2)$ theory. The 
vacuum expectation values for the fileds $S$ and $P^{1 N-2}$ (\ref{vevs}) 
correspond to the expectation values of $\bar{R}_{1 \ldots N-2}$ and 
$\bar{R}_{N}$ of order $v$. 
Correspondingly, due to the first term
in (\ref{wtreebaryon}) 
the fields $Q$ and $\bar{L}^I$ get (supersymmetric) masses of order $\lambda v$.
Since the
$F$ components of the  $\bar{R}$ fields  also have 
expectation values, the fields $Q$ and $\bar{L}^I$ also obtain a supersymmetry
breaking mass squared splitting of order $\lambda F$. 
For the discussion  in the following
section it is relevant to note that the ratio of the 
 supersymmetry breaking mass squared splitting to the supersymmetric mass of the
heavy fields $Q$ and $\bar{L}^I$ is of order $F/v$---the 
same as the corresponding ratio for the light fields in the sigma model.

The components of the $\bar{R}$ fields which get eaten by the $SU(N-2)$ gauge bosons
and their heavy superpartners (with mass of order $g_2 v$) also obtain supersymmetric
mass splitting. 
The leading effect is that
 the scalar components in the heavy vector supermultiplets obtain
supersymmetry breaking contributions to their masses of order $m \simeq F/v$. These 
contributions arise because of a shift of the expectation values of the heavy fields
in response to the F-type vacuum expectation values of the light fields (a similar
effect of the heavy tadpole is discussed in ref.~\cite{BPR}; see also \cite{RP}).

Having understood the supersymmetry breaking vacuum  and the  spectrum 
in some detail we  now turn to using these theories  for model building. 

\section{Communicating Supersymmetry Breaking.}

\subsection{Basic Ideas.}

The basic idea  is to  construct a model  containing two sectors:  the usual
Standard Model sector,  consisting of the  supersymmetric Standard Model 
and  a supersymmetry breaking sector  consisting of  an $SU(N) \times SU(N-2)$
theory  studied above.   We saw above that the latter  theories have an $SP(N-3)$
global symmetry which is left unbroken in the supersymmetry 
 breaking vacuum.  A
subgroup of  $SP(N-3)$   can be    identified with the Standard Model 
gauge symmetries.   
The minimal $SP(2k)$ group in which  one can
embed $SU(3)\times SU(2) \times U(1)$ is  $SP(8)$---this corresponds  to taking
$N=11$.  Alternatively,  we  can consider  an embedding consistent with Grand 
Unification.  For this purpose one can embed $SU(5)$  in $SP(10)$---using the 
$SU(13)\times SU(11)$ models.

The soft parameters---the Standard Model gaugino masses and  
soft scalar masses---receive contributions from several different energy 
scales. As discussed in the previous section, all heavy fields in the 
$SU(N)\times SU(N-2)$ theory that transform under the Standard Model
gauge group obtain supersymmetry breaking mass splittings. The $Q$ and $\bar{L}$ heavy
fields transform as fundamentals under the Standard Model gauge group, whereas
the eaten (and superpartners) components of the fields $\bar{R}$ transform as 
two fundamentals, a symmetric tensor (adjoint), and an antisymmetric tensor 
representation of $SP(N-3)$.

In this section we will present a brief
discussion of the generation of the soft parameters.  As in \cite{dnns}
gaugino masses arise at one loop, while soft scalar masses arise at two
loops. The corresponding calculations are somewhat more involved than the
ones from \cite{dnns}, \cite{martin}; more details 
will be presented in a subsequent paper \cite{future}.

We first consider the  effects  of the heavy $Q$ and $\bar{L}$ fields. 
The contribution of these fields   is analogous to that
of the messenger fields in the models of \cite{dnns}.   Consequently their
contribution to the gaugino masses  is:
 \beq
\label{msoftH}
\delta_H m_{gaugino}  \sim N_f~ {g^2 \over 16 \pi^2}~{F\over v}  \sim
N_f~ {g^2 \over 16 \pi^2}~ M ~ \left( {v \over M} \right)^{N-4} ~,
\eeq  
while their contribution to the soft scalar masses is :
\beq
\label{softscalarH}
\delta_H m_a^2 \sim  N_f ~  {g^4  \over 128 \pi^4} ~ 
C_a ~ S_Q ~ \left(  F \over v \right)^2 ~.
\eeq
In the equations above $g$ denotes the appropriate  Standard Model gauge 
coupling,  $C_a$ is the quadratic Casimir 
($(N^2 - 1)/2 N$ for an $SU(N)$ fundamental; for $U(1)_Y$
the corresponding coefficient is $3 Y^2/5$, with 
$Y$ the messenger hypercharge), $S_Q$ is the Dynkin index of the
messenger representation (1/2 for the fundamental of $SU(N)$).
Finally,  in eqs.~(\ref{msoftH})  and (\ref{softscalarH})  
$N_f$  denotes the number of messenger flavors for
the  appropriate Standard Model groups---in particular it is important to note
that  $N_f$ is proportional to  $N$ so that these contributions increase in
magnitude as the size of the $SU(N) \times SU(N-2)$ group increases.  
 
Next we consider the contributions of the light fields (described by the 
sigma model).  These  effects  will be described in more detail 
elsewhere \cite{future}, here we restrict ourselves to providing some rough 
order of magnitude estimates. Their contribution to the gaugino masses 
is of order:
\beq
\label{msoftL}
\delta_L m_{gaugino}  \sim  {g^2 \over 16 \pi^2} ~ m  \sim 
{g^2 \over 16 \pi^2}~ M ~ \left( {v \over M} \right)^{N-4} ~,
\eeq  
where $m$  denotes the typical  mass scale in the sigma model,
 eq.~(\ref{orderofmasses}).
Since 
 the supertrace
of the light messenger mass matrix squared is nonvanishing (as can be inferred
from eqs.~(\ref{diracmass}), (\ref{scalarmass}), and Table 1), their
contribution to the soft scalar masses turns out to be logarithmically 
 divergent \cite{future}. The divergent piece is:
\beq
\label{softscalarL}
\delta_L m_a^2 = - ~ {g_a^4  \over 128 \pi^4} ~ C_a ~ S_Q ~
{\rm Str} M^2_{mess} ~{\rm Log} {\Lambda^2\over m_f^2} ,
\eeq
where $\Lambda$
 is the ultraviolet
cutoff and $m_f $ is the Dirac mass of the messenger fermion, 
$m_f \sim F/v\sim m$.
This logarithm is cutoff by the contributions of the
 heavy eaten components of
the fields\footnote{As far as the soft Standard Model parameters are
concerned, this is the main effect of the  supersymmetry breaking mass
splittings in the  eaten components of the $\bar{R}$ fields. } $\bar{R}$,
therefore the scale $\Lambda$ in  eq.~(\ref{softscalarL}) should be replaced by
their mass,  $\sim g_2 v$  \footnote{In the full theory, the ${\rm Str} M^2$,
appropriately weighted by the messengers' Dynkin indices vanishes. One can
see this by noting that a nonvanishing supertrace would imply the existence
of a counterterm for the Standard Model soft scalar masses. This counterterm
would have to be nonpolynomial in the fields (for example, of the form 
$\Phi^\dagger \Phi {\rm Log} \bar{R}^\dagger \bar{R}$) 
and thus can not appear.}.  Note
also that in (\ref{softscalarL})  ${\rm Str} M^2_{mess} \sim m^2$, and  that
there is no large flavor factor $N_f$, since there are only two fundamentals of
$SP(N-3)$ light messengers. In addition to the logarithmically divergent
contribution (\ref{softscalarL}),  there are finite contributions analogous to
those of the heavy fields (\ref{softscalarH}),  which   are proportional to
$\Delta m_{mess}^2/m_{mess} \sim F/v \sim m$  (the exact formula will be
given in \cite{future}).\footnote{For 
completeness, we note that with the general messenger
scalar mass matrix (\ref{scalarmass}), one loop contributions to
the hypercharge D-term are generated. These can be avoided if
the messengers fall in complete $SU(5)$ representations, or,
alternatively, the parameter $a$ is sufficiently large (for $a$ not sufficiently
large, however, there are two loop 
contributions to the $U(1)_Y$ D term,
even in the complete $SU(5)$ representations case).}

We can now use the above estimates for the Standard Model
soft masses, eqs.~(\ref{msoftH}), (\ref{softscalarH}), (\ref{msoftL}),  
and (\ref{softscalarL}), to obtain
an estimate of the scales in the $SU(N)\times SU(N-2)$ theory.
In section 2.3.1, we found that the scale of the messenger
masses (\ref{orderofmasses})
 is given by  $m = M (v/M)^{(N-4)} = F/v$, while the scale of supersymmetry breaking
(\ref{ftermscale}) is $\sqrt{F} = M (v/M)^{(N-3)/2}$. 
Demanding, e.g. that $m_{gaugino} \sim (10^2 - 10^3)$ GeV, we obtain
\beq
\label{voverM}
{v \over M} \sim \left({ (10^4 - 10^5)~{\rm GeV} \over M} \right)^{1\over N-4}~.
\eeq
The scale of supersymmetry breaking  (\ref{ftermscale}) then becomes
\beq
\label{susybreakingscale}
\sqrt{F} \sim M ~\left({ (10^4 - 10^5) ~{\rm GeV} \over M} 
\right)^{N-3 \over 2(N-4)}~.
\eeq

\subsection{Hybrid Models.}

Since $M$ suppresses the non-renormalizable operators in eq. (\ref{wtreebaryon})
one natural  value it can take is  $M_{Planck}$.  We consider this case in some 
detail  here.  On 
setting $M = M_{Planck} \simeq 2 \cdot 10^{18}$ GeV in the formula above gives
$\sqrt{F} \sim 10^{18} (10^{-14}-10^{-13})^{ N-3 \over 2(N-4)}$ GeV. 
As discused above, the smallest value of $N$  for which the Standard Model groups can be 
embedded in the flavor group is $N=11$. This corresponds to 
$\sqrt{F} \sim 10^{10}$  GeV, i.e.  the supersymmetry
breaking scale is of order the intermediate scale. 
It also follows from eq.~(\ref{susybreakingscale}) that on increasing $N$,
the scale of supersymmetry breaking increases very slowly. 
For example, with $N=13$---the smallest value 
consistent with Grand Unification---$\sqrt{F} \sim 10^{10} - 10^{11}$ GeV, 
still of order the intermediate scale.
One consequence of the supersymmetry breaking scale being of order the 
intermediate scale is that 
the squark and slepton masses due to supergravity, of order $F/M_{Planck}$,
 will be comparable to the  
masses induced by the gauge interactions. 
 These models can therefore be 
thought of as ``hybrid models" in which scalar masses arise
 due to both supergravity and gauge interactions,
while gaugino masses arise solely from the gauge interactions.

It is also illustrative to work out the other energy scales in the supersymmetry 
breaking sector. For concreteness we focus on the $N=11$ theory. 
{}From eq.~(\ref{vevscale}) 
we find that $v \sim 10^{16} $ GeV while  from eq.~(\ref{approx}),  it follows that
$\Lambda_{1L} \sim 10^{12}$ GeV.  Notice in particular that $\Lambda_{1L} 
\ll v$ so that the requirement in eq.~(\ref{approx}) is met and  the approximations
leading to the sigma model are valid. The
underlying physics giving rise  to supersymmetry breaking in this model can  
be described as follows. One starts  with a $SU(11) \times SU(9)$ theory at very high
energies.  At $v \sim 10^{16} $ GeV,  the $SU(9)$ symmetry is broken giving rise to 
a theory consisting of some moduli and  a pure $SU(11)$ group coupled to a 
dilaton. The $SU(11)$ group confines at $\Lambda_{1L} \sim 10^{12}$ GeV,
 giving rise to a sigma model consisting of the moduli and the dilaton. 
Finally, supersymmetry 
breaks at $10^{10}$ GeV giving rise to masses for 
messenger quarks of order $10$ TeV. 
It is worth noting that this large hierarchy of scales is generated dynamically.    
We also note that this hybrid model does 
not exhibit Landau poles (below scales, higher than $v \sim 10^{16}$ GeV) 
 of the Standard Model gauge groups: between
the messenger scale and the scale $v$, 
 in addition to the usual quark, lepton
and Higgs supermultiplets only two vectorlike $SU(3)$ flavors  and two 
$SU(2)$ fundamentals contribute to the running of the gauge couplings. Above
the scale $10^{16}$ GeV, new physics is expected to take over, as discussed in the
Introduction.

The high scale of supersymmetry breaking in these models poses a problem and 
constitutes their  most serious drawback.
It implies that one cannot generically rule out the presence of large flavor 
changing neutral current effects.  Such effects could arise   
due to higher dimensional operators in the K\" ahler potential. 
For these models  to be viable, 
physics at the Planck scale would have to prevent such operators from 
appearing. In this respect these models are no better than the usual hidden 
sector models.

It is worth emphasizing  the  key features  of the $SU(N) \times SU(N-2)$
theories that are  ultimately responsible for the high scale of 
supersymmetry  breaking.  The requirement that the flavor group is big enough
forces  one to large 
values of $N$ in these theories\footnote{For smaller values 
of N,  $N\le 7$, the scale of supersymmetry breaking
$\sqrt{F} \le 10^9$ GeV, and the problem of flavor changing effects may
be alleviated. However, in this case, 
we can not embed the whole Standard Model gauge group in
the unbroken $SP(N-3 \le 4)$ global symmetry (in particular, the gluinos
would have to be massless in this framework).}. Furthermore, supersymmetry 
breaking occurs only in the presence  of nonrenormalizable 
operators whose dimension grows with $N$.   Suppressing these 
operators by the Planck scale leads to the high scale of supersymmetry 
breaking. 

\subsection{Purely Gauge Mediated Models. }

One would like to find  other theories  in  which  the  requirement for 
a big enough flavor symmetry can be met without leading to such a 
high supersymmetry breaking scale.  We discuss two possibilities
in this context. 

\subsubsection{Lowering the scale $M$.}

One possible way in which the supersymmetry breaking scale can 
be lowered is by making $M < M_{Planck}$.
The $SU(N) \times SU(N-2)$ theory of Section 2.1  itself would in this case be 
 an effective 
theory,  which would arise from some underlying dynamics at scale $M$. However, 
to suppress the flavor changing neutral currents one would have to 
forbid $D$ terms of the form 
\beq
\label{kahlerterms}
{\bar{R}^\dagger~ \bar{R}~ \Phi^\dagger ~\Phi\over M^2}~,
\eeq
where $\Phi$ denote Standard Model fields, in the effective theory. 
Such terms, if present in a flavor
 non-universal form, would be problematic (at least for $N$ sufficiently 
large to accommodate the whole Standard Model gauge group).
It is possible that they might be absent in a theory where the last two 
terms in eq.~(\ref{elsup}) arose 
due to non-perturbative dynamics that only couples to the $\bar{R}$
fields but not to the Standard Model.   

Once the supersymmetry breaking scale is lowered these  theories 
can be used to construct purely gauge mediated models of supersymmetry 
breaking. 
The feeddown of supersymmetry breaking to the Standard Model in these 
models 
proceeds as described in Section 3.1. Both gaugino  and scalar soft masses
receive contributions 
from  the heavy, eqs.~(\ref{msoftH}), (\ref{softscalarH}),
 and light, eqs.~(\ref{msoftL}), (\ref{softscalarL}), messengers.  
As follows from eq.~(\ref{softscalarL}) and the sigma model spectrum of
Table 1 (note that ${\rm Str} M^2_{mess} \sim \alpha + \beta$), 
 the logarithmically enhanced contribution of
the light fields to the scalar masses  is in fact negative.  Consequently,
obtaining positive soft scalar mass squares
 poses a significant constraint on the models. 
These masses can be positive if the additional finite contributions of
the heavy and light messengers overcome the negative 
logarithmically enhanced contribution
of the light messengers.
This can happen in two ways. 
First the logarithmic contribution can be reduced in magnitude by lowering 
the scale $g_2 v$, which cuts off the logarithm, and bringing it sufficiently
close to the scale $m$.  For example, with $N=11$,
using Table 1, one can conclude that positive mass squares are obtained with a 
scale $v$ two orders of magnitude larger than the scale $m \sim 10^{4} - 10^5$ GeV.
Note that lowering the
scale $g_2 v$ 
amounts to lowering the scale $M$, eq.~(\ref{wtreebaryon}), at which new
physics must enter.
 Second, we note that 
the positive finite contributions, 
eq.~(\ref{softscalarH}), 
of the heavy fields $Q$ and $\bar{L}$, 
are enhanced by a factor of $N_f  \sim  N$. In addition, as is clear from 
the numerical results of Table 1, with 
increasing $N$ the ratio of the supertrace (proportional to
$\alpha + \beta$) to the finite contribution (proportional to $(\delta/\gamma)^2$)
 decreases. Consequently,  models with 
$N$ sufficiently large 
will yield positive mass squares, without requiring the scale $M$ to be too close
to the scale of the light messengers\footnote{Similar observations have been made
recently in ref.~\cite{berkeley}. We thank J. March-Russell for discussions in this
regard.}. 
Having the scale $M$ be as large as the 
GUT scale pushes the Landau poles up, which is an attractive feature of the
models that one might want to retain. We leave a detailed analysis of this issue
for future work \cite{future}. We only mention here 
the phenomenologically
interesting possibility that the two competing effects, (\ref{softscalarH}), 
and (\ref{softscalarL}) might yield squarks that 
are lighter than the gauginos.

We conclude this section by raising  the possibility that 
the scale  $M$ could be less than  $ M_{Planck}$ if the 
Standard Model gauge 
groups are dual to some underlying theory.
 In order to illustrate this, we return to our starting point,
the $SU(N) \times SU(N-2)$ theory, with,  as discussed above,
 the Standard Model groups embedded in the $SP(N-3)$ global symmetry.  
As a result of the additional degrees of freedom the 
Standard Model  groups are severely non-asymptotically free, 
once all the underlying
degrees in the $SU(N) \times SU(N-2)$ theory come into play. Consequently,
it is appealing to dualize the theory and to regard the dual, which is 
better behaved in the ultraviolet, as the underlying microscopic theory. 
  We see below that this could also lead to lowering the scale $M$,  
eq.~(\ref{wtreebaryon}), in the electric theory. 

For purposes of illustration we work with the $N=11$ case and consider 
dualizing the Standard Model $SU(3)$ and $SU(2)$ groups. 
In the process we need to 
re-express the baryonic operators in eq. (\ref{wtreebaryon}) in terms of 
gauge invariants 
of the two groups and then use  the duality transformation of 
SQCD, \cite{seiberg},  to map 
these operators to  the dual theory. Doing so shows that 
the baryonic operators can be expressed  as a product involving some 
fields neutral under the Standard Model  groups and mesons of the 
$SU(3)$ and $SU(2)$ 
groups. But the mesons  map to fields which are singlets 
in the dual theory. 
Consequently, the resulting terms in the 
superpotential of the dual theory have smaller canonical dimensions 
and are therefore suppressed by fewer powers of $M_{Pl}$. 
For example, the operator $b^{N-1 N-2}$ can be written as a product involving
the field $\bar{R}_N$, three mesons of $SU(3)$, and a meson of $SU(2)$; 
as a result in the dual it has dimension $5$ and is suppressed by 
two powers of $M_{Pl}$. The deficit in terms of dimensions is made up 
by the scales $\mu_3$ and $\mu_2$ which enter the scale matching relations
for the $SU(3)$ and $SU(2)$ theories, respectively \cite{seiberg}, 
 leading to a relation:
\beq
\label{scale matching}
M = M_{Pl}~ \left ({\mu_3^3 ~\mu_2 \over M_{Pl}^4 }\right )^{1 \over 6}.
\eeq
For $\mu_3$ and $\mu_2$ much less than $M_{Pl}$ we see that $M$ is 
much lower than $M_{Pl}$.

While the above discussion is suggestive,  several concerns need to 
be met before it can be made more concrete. First,
as was mentioned above, one needs to argue that terms of the form 
 (\ref{kahlerterms}) are suppressed 
adequately. We cannot at present, conclusively,   settle this matter since the 
map for non-chiral operators under duality is not known. However, since
the scales involved in the duality transformation are much smaller than 
the Planck scale,  it is quite plausible that if an operator of the form 
eq. (\ref{kahlerterms}), suppressed by the Planck scale,   is present in the 
dual theory it will be mapped to an operator in the electric theory 
that is adequately suppressed.  
Second, in the example discussed above, the Standard Model $U(1)_Y$ 
group continues to be 
non-asymptotically free. This can be avoided by 
considering  theories in which the Standard Model groups are embedded in a 
GUT group. The simplest such
example is the  $N=13$ theory with a GUT group $SU(5)$. The $SU(5)$ group
has matter in the fundamental, antisymmetric and adjoint
 representations. Unfortunately, no compelling dual for this theory
is known at present. \footnote{This theory
 can be dualized by following the methods of \cite{Pouliotwo}, \cite{Kutasov}
 and
unbinding each antisymmetric tensor by introducing an extra $SU(2)$ group.
However, the resulting dual is quite complicated and contrived.}
Finally, the above attempt at lowering 
$M$ relied on taking the  parameter(s) $\mu$ to be  smaller than 
$M_{Pl}$. This might be unnatural in the dual theory.
 For example, in the dual theory considered here, 
the Yukawa coupling $\lambda^{IA} Y_{IA}$, eq. (\ref{wtreebaryon}),  turns 
into a mass term with a $\mu$ dependent coefficient.  Naturalness, in 
this case suggests that $\mu$ is of order $M_{Pl}$.
A detailed discussion of these issues is left for the  future, hopefully, within 
the context of more compelling models and their duals.

\subsubsection{Other Sigma Models.}

We saw  in our discussion of the hybrid models  above that a large hierarchy
of scales   separates the microscopic
theory from the sigma model. In view of this, one can ask if at  least a sigma
model can be  constructed as an effective theory that yields  a low enough
supersymmetry breaking scale, while the nonrenormalizable operators are
still suppressed by the Planck scale. The answer, it is easy to see, is yes.
For example,  we can take the dimensions of the fields in the effective
lagrangian (\ref{effectivelagrangiankah}), (\ref{elsup}) to be equal to, say,  
$D$---being thus different 
from their dimension, $N-2$,  dictated by
the underlying $SU(N)\times SU(N-2)$ theory---and change
correspondingly the power of the $1/M$-factors, the powers in the K\" ahler
potential and the power of $S$ in the
nonperturbative term in the superpotential,  eq.~(\ref{elsup}).
We should emphasise that we are not aware of any underlying microscopic
theory which gives rise to such a sigma model. However,  they do  
provide an adequate description of supersymmetry breaking. 
An analysis similar to the one  above shows that these sigma models
break supersymmetry, while leaving an $SP(N-3)$ flavor subgroup
intact. The mass spectrum of low lying excitations in these theories
is also qualitatively of the form in eq.~(\ref{scalarmass}) and 
eq.~(\ref{diracmass}).  
Following then the same arguments that lead to eq.~(\ref{susybreakingscale})
for the supersymmetry breaking scale, we find that the exponent
in eq.~(\ref{susybreakingscale}) changes to $(D-1)/(2(D-2))$ instead.
Consequently, for $D = 4 $ or $5$ (even with $M = M_{Planck}$),
the scale of supersymmetry breaking is sufficiently low for supergravity
effects to be unimportant. 

It is illustrative to compare the energy scales obtained in such a model 
with those obtained in the ``hybrid" models above. We consider the $D=4$
case for concreteness. The supersymmetry breaking scale in this case is 
of order $10^7$ GeV, well below the intermediate scale,
while the scale of the vacuum 
expectation values is $\sim 10^{11}$ GeV. Therefore the 
the sigma model breaks down  at an energy scale well above the scale of 
supersymmetry breaking.\footnote{The scale at which the effective theory
breaks down could be smaller than the perturbative estimate 
coming from the sigma model, $\sim 4 \pi v$, would indicate. For example, 
if we had retained the corrections to the K\" ahler potential of
order $\Lambda_L/v$, discussed in Section 2.2.1, we would have found that
the model breaks down at a scale $\Lambda_L$, which is 
lower than $v$, but still higher than $M_{SUSY}$.}

Once the supersymmetry breaking scale is sufficiently lowered  one 
can use these sigma models to construct purely gauge mediated models of
supersymmetry breaking.  We note, however, that we can not 
compute the Standard Model soft masses from the effective theory 
alone---we saw in 
Section 3.1 that the contribution of the  heavy states not included in the
sigma model  can be as
important as the ones from the light fields.

\section{Phenomenological Implications.}

In this section, we discuss 
the phenomenological implications
of the ``hybrid" models of dynamical supersymmetry breaking, introduced
above.  Towards the end we will briefly comment on some expected features  
of purely gauge mediated models  with a combined 
supersymmetry breaking and messenger sector. 
In our discussion of  hybrid models  we will,  where  necessary, 
focus on the  $SU(11)\times SU(9)$ model, 
in which the    $SU(3)\times SU(2) \times U(1)$ groups are embedded 
in the $SP(8)$ global symmetry group.
 
We begin with two observations. First, since the supersymmetry breaking 
scale is high in these models, the gravitino has a weak scale mass and 
is not (for non-astrophysical purposes at any rate) the LSP.
Second, since the supersymmetry breaking sector is coupled quite directly
to the Standard Model sector, the masses of the (light) fields in the 
supersymmetry breaking
sector are of order $10$ TeV. Consequently, at this scale one can 
probe all the fields that play an essential role in  the breaking of supersymmetry. 

We now turn to the scalar soft masses. As noted in the previous section,
scalars in these models receive contributions due to both gauge and 
gravitational effects. Gravitational effects give rise 
to  universal soft masses of
order $F/M_{Planck} \sim 10^2 - 10^3$ GeV at the Planck scale\footnote{
We are assuming as usual here that the K\" ahler metric is flat.}. 
In addition, as described in Section 3.1, Standard Model gauge interactions 
induce non-universal contributions (\ref{softscalarL}), (\ref{softscalarH}). 
Since the soft masses receive contribution at various energy scales, 
the  renormalization group running in the hybrid models is quite different from the running
in  supergravity hidden sector models and from that in gauge mediated, low-energy 
supersymmetry breaking models. 
We leave the detailed study of the renormalization group effects for future work.

Getting a big enough $\mu $ term in these models is a problem.  
Since the model is ``hybrid", one could attempt to use  $1/M_{Planck}^2$--suppressed 
couplings, such as 
$\int d^4 \theta H_1 H_2 \bar{R}^{\dagger} \bar{R}$ 
or $\int d^2 \theta H_1 H_2 (W^\alpha W_\alpha)_{SU(N)}$, 
to generate  the desired  $\mu$ and $B \mu$ terms. However, it is easy to see that 
while $B \mu \sim F^2/M_{Planck}^2$ is generally of the right order of 
magnitude, 
the resulting $\mu$-parameter is 
$\mu \sim (v/M_{Planck}) \sqrt{B \mu} \sim 10^{-2} \sqrt{B \mu}$ and is therefore too small.
A similar conclusion results from considering, e.g. the $F$ term $b H_1 H_2/M^{N-3}$,
with $b$ being an $SP(N-3)$-singlet baryon, 
which can be used to generate a reasonable  $B \mu$ term and a negligible $\mu$ term
(to see this, we use 
$\langle b/M^{N-3} \rangle \sim M (v/M)^{N-2} + \theta^2 M^2 (v/M)^{2(N-3)}$, with
$M \sim 10^{18}$ GeV, $v \sim 10^{16}$ GeV, and $N=11$). 

To avoid this small-$\mu$ problem, 
one could   use the  approach of  ref.~\cite{dnns} and introduce a special sector 
of the 
theory, constrained by some discrete symmetry, 
 which will be responsible for generating the $\mu$ term.  For example, this 
 could be achieved by requiring an  appropriate $SP(N-3)$-singlet baryon, 
$b/M^{N-3}$,  to play the role  of  the singlet field  $S$ of ref.~\cite{dnns}  
(see Section 4 of last paper in \cite{dnns})  
and the introduction of  an additional singlet $T$ with appropriate 
couplings in the superpotential.   
{}From the point of view of low-energy phenomenology, 
this approach implies that when analyzing  the
low-energy predictions of the model, $\mu$ and $B \mu$  should  be treated as
free parameters.

A few more comments are in  order. 

First,   electroweak symmetry breaking  will  occur  radiatively  in these models,
with the large top Yukawa driving the mass  square of one Higgs field negative. 
Second,  these models  do not suffer from a  supersymmetric CP problem. 
This can be seen immediately in the sigma model superpotential eq.~(\ref{elsup}), 
where all  phases can be  rotated away.\footnote{It can also be seen in the 
underlying  $SU(N) \times SU(N-2)$ theory where all phases except for the $\theta$
angle of $SU(N-2)$ can be rotated away.  Since the $SU(N-2)$ group is broken at 
a very high scale,  its instantons are highly suppressed.} 
Finally, we note that the hybrid models  are likely  to inherit   some of the 
cosmological  problems of hidden sector models.  For  example, the $R$ axion, whose
mass in this model can be seen to be of order the electroweak scale \cite{BPR}, 
is very weakly interacting, $f_{axion} \sim v \sim 10^{16}$ GeV, and may suffer
the usual Polonyi problem.  This problem could be  solved,   for example,  by  
invoking  weak scale inflation.  

We end with a few comments about the phenomenological implications  of 
purely gauge mediated models with  a
supersymmetry breaking-cum-messenger sector.   
As was mentioned in Section 3,
such models can be constructed by lowering the scale $M$.
 A few key  features
emerge from considering  such purely gauge mediated
 models, which are likely to be generally true 
in  models   of this kind.  
 First, as we have seen above, 
the scale of supersymmetry 
breaking  which  governs the mass and interaction strength  of the  gravitino,
is a parameter which can take   values   ranging from  $10$ TeV to $10^{10}$ GeV 
and can  therefore be very different from the  value  of  the 
messenger  field masses.  It should therefore
be treated as an independent parameter in considering the phenomenology of these 
models.  
Second,  one consequence of having a combined 
supersymmetry breaking and messenger sector is that  several 
  degrees of freedom responsible for  the communication and 
the breaking of supersymmetry breaking can be probed at 
 an energy of about $10$ TeV.
 Finally, the  form of the mass matrix of the messenger fields can be different from 
that in the models of ref.~\cite{dnns}, as is clear from  eqs.~(\ref{scalarmass}) and
(\ref{diracmass}).   In  particular, the sum rule relating the fermion and boson
masses is not respected in general.  We expect this to be a general feature of such
models. As discussed in Section 3.1, the nonvanishing supertrace for the light
messenger fields gives a logarithmically enhanced contribution to the soft
scalar masses. In the models discussed here, the supertrace is positive and
the corresponding contribution to the soft scalar masses squared is negative.
This poses a constraint on model building. The negative contribution can be 
controlled by lowering the scale $M$, or considering models with large $N$.
This could lead to scalar soft masses that are lighter than 
the gaugino masses\footnote{
We acknowledge discussions with G. Anderson on this point.}.
A detailed analysis of the spectrum 
and the resulting phenomenology 
is left for the future \cite{future}.

\section{Summary.}

In conclusion we  summarize the main results of this paper and indicate
some possible areas for  future study:
\begin{itemize}

\item{
We began this   paper  by studying a class of supersymmetry
 breaking theories  with an  $SU(N) \times SU(N-2)$  gauge group.  
We showed how   the breaking of supersymmetry
 in   these   theories  can  be studied  in a  {\it calculable} 
low-energy sigma model.   The sigma model   was used to show   that a large 
subgroup of  the global symmetries    is  left unbroken  in these theories, 
 and   to calculate the  low-energy mass spectrum  after supersymmetry breaking. } 

\item{ 
We then turned  to using these theories   for  model building.   The models
we constructed had two sectors: a supersymmetry breaking sector,  consisting of the 
above mentioned $SU(N) \times SU(N-2)$ theories, and  the 
supersymmetric Standard Model.  The essential idea 
was to identify  a subgroup of  the  global symmetries  
of the  supersymmetry breaking sector with  the Standard Model gauge group.  
In order to embed the full  Standard Model gauge group  in this way, we were 
lead  to  consider   large values of $N$, i.e. $N \ge 11$,  and  as   a consequence
 of this large value of $N$, the supersymmetry
 breaking scale  was driven up to be of order the
intermediate scale, i.e. $10^{10}$ GeV.   Hence, these models   
are of a ``hybrid" kind---supersymmetry 
breaking is communicated  to the Standard Model  both
gravitationally and  radiatively through the Standard Model gauge groups 
in them.}

\item{We briefly discussed the phenomenology  of these models.
The main consequence of  the  messenger fields being an integral 
part of the supersymmetry
 breaking sector
is that several  degrees of freedom responsible for  both communicating and 
breaking  supersymmetry  can be probed at an energy of order $10 $ TeV. 
In the hybrid models  gauginos acquire mass due to gauge mediated effects, while  scalars
acquire mass due to both gauge and gravitational effects.  We leave  a more
detailed investigation of  the resulting  mass spectrum, including the effects 
of renormalization group running for further study.} 

\item{ 
It is worth mentioning that  in these models there is a  large hierarchy of
scales that is generated  dynamically. For example, even though
the scale of supersymmetry breaking  is high, of order $10^{10}$ GeV, 
  the masses of the messenger fields---the lightest 
 fields in the supersymmetry
 breaking sector  that carry Standard Model
charges---are of order $10$ TeV.  Furthermore, 
the sigma model used for studying the low-energy dynamics 
breaks  down at a scale  $10^{12}$ GeV---well above  the scale of
supersymmetry breaking. }

\item{Purely gauge mediated models can be constructed by lowering the
scale $M$ that suppresses the nonrenormalizable term in the superpotential.
These  purely gauge mediated models  reveal  the following   features that
should  be  generally true in models  with   supersymmetry
breaking-cum-messenger sector that have an effective low-energy
weakly coupled  description. 
 First,  the supersymmetry
 breaking scale  can  in general
be quite different from the scale of the messenger field masses---it can range from 
$10$ TeV to $10^{10}$ GeV, 
while the messenger field masses are of order $10$ TeV.  
Second, as in the hybrid models, several degrees of freedom that
are responsible for  communicating and breaking  supersymmetry can be probed 
at an energy scale or order $10$ TeV.
Third, 
the Standard Model 
soft masses receive contributions at various energy scales.  
Because of a tradeoff between positive and negative contributions, 
the soft scalar masses can be lighter than
the corresponding gaugino masses.
  A detailed  investigation of the 
phenomenology of such models,   incorporating   these
features,  needs to be  carried out.    We leave such an  investigation for the
future.}  

\item{Finally, we hope to return to the construction of purely gauge mediated 
models of supersymmetry breaking with a 
combined supersymmetry breaking and messenger 
sector.  One  would like to construct a  consistent microscopic theory
which could give rise to an  adequate supersymmetry breaking sector. 
A minimal model  of this kind  would serve to further guide
phenomenology. It would also  prompt an investigation of  more theoretical
questions---like those associated with the loss of asymptotic freedom for the
Standard Model gauge groups.}

\end{itemize}

We would like to acknowledge discussions with G. Anderson, J. Lykken, 
J. March-Russell, S. Martin, 
and especially Y. Shadmi. Recently, we became aware of work by 
N. Arkani-Hamed, J. March-Russell, and H. Murayama along similar 
lines \cite{berkeley},
and thank them for sharing some of their results before publication.
E.P. acknowledges support by a Robert R. McCormick
Fellowship and by DOE contract DF-FGP2-90ER40560. 
 S.T. acknowledges 
the support of DOE contract DE-AC02-76CH0300.

\nc{\ib}[3]{ {\em ibid. }{\bf #1} (19#2) #3}
\nc{\np}[3]{ {\em Nucl.\ Phys. }{\bf #1} (19#2) #3}
\nc{\pl}[3]{ {\em Phys.\ Lett. }{\bf #1} (19#2) #3}
\nc{\pr}[3]{ {\em Phys.\ Rev. }{\bf #1} (19#2) #3}
\nc{\prep}[3]{ {\em Phys.\ Rep. }{\bf #1} (19#2) #3}
\nc{\prl}[3]{ {\em Phys.\ Rev.\ Lett. }{\bf #1} (19#2) #3}
\nc{\ptp}[3]{ {\em Progr.\ Theor.\ Phys.}{\bf #1} (19#2) #23}

\end{document}